\documentclass[final,english]{bullsrsl}[2022/06/15]

\usepackage[latin1]{inputenc}
\usepackage[T1]{fontenc}

\usepackage{natbib} 
\usepackage{graphicx}

\begin{document}
\title{Automated transient detection in the context of the 4m ILMT}

\author[affil={1,2}, corresponding]{Kumar}{Pranshu}
\author[affil={1,3}]{Bhavya}{Ailawadhi}
\author[affil={4,5}]{Talat}{Akhunov}
\author[affil={6}]{Ermanno}{Borra}
\author[affil={1,7}]{Monalisa}{Dubey}
\author[affil={1,7}]{Naveen}{Dukiya}
\author[affil={8}]{Jiuyang}{Fu}
\author[affil={8}]{Baldeep}{Grewal}
\author[affil={8}]{Paul}{Hickson}
\author[affil={1}]{Brajesh}{Kumar}
\author[affil={1}]{Kuntal}{Misra}
\author[affil={1,3}]{Vibhore}{Negi}
\author[affil={8}]{Ethen}{Sun}
\author[affil={9}]{Jean}{Surdej}
\affiliation[1]{Aryabhatta Research Institute of Observational sciencES (ARIES), Manora Peak, Nainital, 263001, India}
\affiliation[2]{Department of Applied Optics and Photonics, University of Calcutta, Kolkata, 700106, India}
\affiliation[3]{Department of Physics, Deen Dayal Upadhyaya Gorakhpur University, Gorakhpur, 273009, India}
\affiliation[4]{National University of Uzbekistan, Department of Astronomy and Astrophysics, 100174 Tashkent, Uzbekistan}
\affiliation[5]{ Ulugh Beg Astronomical Institute of the Uzbek Academy of Sciences, Astronomicheskaya 33, 100052 Tashkent, Uzbekistan}
\affiliation[6]{Department of Physics, Universit\'{e} Laval, 2325, rue de l'Universit\'{e}, Qu\'{e}bec, G1V 0A6, Canada}
\affiliation[7]{Department of Applied Physics, Mahatma Jyotiba Phule Rohilkhand University, Bareilly, 243006, India}
\affiliation[8]{Department of Physics and Astronomy, University of British Columbia, 6224 Agricultural Road, Vancouver, BC V6T 1Z1, Canada}
\affiliation[9]{Institute of Astrophysics and Geophysics, University of Li\`{e}ge, All\'{e}e du 6 Ao$\hat{\rm u}$t 19c, 4000 Li\`{e}ge, Belgium}

\correspondance{pranshu@aries.res.in}
\date{15th May 2022}
\maketitle


%

\begin{abstract}
In the era of sky surveys like Palomar Transient Factory (PTF), Zwicky Transient Facility (ZTF) and the upcoming Vera Rubin Observatory (VRO) and ILMT, a plethora of image data will be available. ZTF scans the sky with a field of view of 48 deg\textsuperscript{2} and VRO will have a FoV of 9.6 deg\textsuperscript{2} but with a much larger aperture. The 4m ILMT covers a 22$'$ wide strip of the sky. Being a zenith telescope, ILMT has several advantages like low observation air mass, best image quality, minimum light pollution and no pointing time loss. Transient detection requires all these imaging data to be processed through a Difference Imaging Algorithm (DIA) followed by subsequent identification and classification of transients. The ILMT is also expected to discover several known and unknown astrophysical objects including transients. Here, we propose a pipeline with an image subtraction algorithm and a convolutional neural network (CNN) based automated transient discovery and classification system. The pipeline was tested on ILMT data and the transients as well as variable candidates were recovered and classified.
\end{abstract}


\section{Introduction}
\label{introduction}

The 4m International Liquid Mirror Telescope (ILMT) achieved first light on 29$^{\rm th}$ April, 2022 \citep{article_firstlight} and is at present in the commissioning phase. The ILMT will survey the zenith sky for a minimum period of 5 years. One major science goal of the ILMT is to search for transients like supernovae \citep{10.1093/mnras/sty298}. Some of these transients are very rare and often require automated techniques to find them. Large all-sky surveys like Zwicky Transient Facility (ZTF) use machine learning-based techniques \citep{Mahabal_2019} to assist in transient discovery. 
    
The ILMT is acquiring a huge volume ($\sim$15 Gigabytes) of data each night. Each ILMT frame can contain a large number of sources and to find a few transients in such an image is like finding a needle in a haystack. The other challenge is to find transients in near real-time. This is because a lot can be understood about the nature of the transients by discovering them in their early stages. It was therefore required to develop an automated pipeline to detect transients occurring in the ILMT strip to fulfil the aforementioned requirements.

The pipeline is based on image-based real/bogus classifier technique. It uses Convolutional Neural Network (CNN) \citep{726791} to carry out the transient detection. It also houses a transient classifier in the pipeline to carry out the classification of the detected sources into three categories {\it viz.} orphan candidates, variable source candidates and extragalactic candidates. The proposed pipeline is described in Section\,\ref{pipeline}. The initial results from the pipeline are given in Section\,\ref{results}. A brief summary and conclusions of this work are presented in Section\,\ref{summary}.

\section{Pipeline Overview}
\label{pipeline}
The pipeline contains three main steps - i) The first step performs image differencing by subtracting reference images from new science images, ii) the second step searches for transients in the resulting difference image and iii) the third step classifies the detected transients into three classes {\it viz.} orphan candidates (e.g. asteroid), variable source candidates (e.g. variable star) and extragalactic candidates (e.g. supernovae, TDEs, AGNs).

\begin{figure}[t]
\centering
\includegraphics[width=\textwidth]{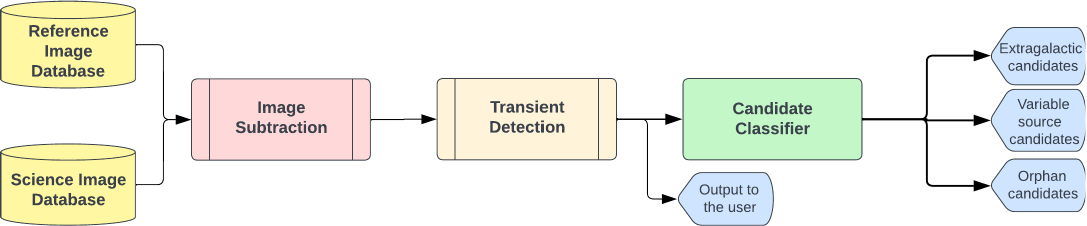}
\smallskip

\begin{minipage}[c]{12cm}
\caption{Schematic diagram of the transient detection pipeline for the 4m ILMT}
\label{PyLMT}
\end{minipage}
\end{figure}

The first two steps of the transient detection pipeline along with the third classifier step offers extragalactic candidates (possibly supernovae) for follow-up in real-time. It also outputs a list of other transients like orphan and variable source candidates after each night of observation. These objects can be followed up subsequently for further photometric/spectroscopic analysis. The various steps involved in the pipeline have been developed separately. All these steps have been integrated into an end-to-end transient detection pipeline. Fig.\,\ref{PyLMT} illustrates the basic blocks of the transient detection pipeline. The three primary steps of the pipeline are elaborated below.

\textbf{Image Differencing:} The first step in the pipeline is the algorithm to perform difference imaging. This technique allows for the detection of transient candidates or varying sources across a pair of images \citep{Alard_1998,Bramich_2008}. It subtracts a reference image with a better seeing \citep{10.1093/mnras/stu835} from the newly acquired science image of the same region in the ILMT field. In image differencing, the science and reference images are aligned using \texttt{astroalign} \citep{astroalign}, space varying 2D median background is removed using \texttt{photutils} \citep{2016ascl.soft09011B}, PSFs of science and reference images are matched using \texttt{scipy} optimizer \citep{Virtanen_2020} and finally the image subtraction is performed. The resultant difference images contain several artefacts as well as real sources. Implementing a simple source finder algorithm will result in the detection of artefacts along with real sources which need to be distinguished by the transient detection step. 

\textbf{Transient Detection:} The second step in the pipeline performs the transient detection. The CNN-based real/bogus classifier \citep{7727206,10.1093/mnras/sty613,Duev_2019} is the core of this step. The CNN-based real/bogus classifier works by accepting 31$\times$31 pixels cutout images of the sources (both real and artefacts) in the difference images with peak intensity above the minimum detection threshold and by giving a class probability value corresponding to each of these sources. The final classification of the sources as \textit{real} or \textit{bogus} (artefact) will depend upon the class probability value and classification threshold (which was kept at 0.5). Apart from the real/bogus classifiers, threshold cuts on parameters like FWHM and peak values of candidate sources have been used.  

The \textit{real/bogus classifier} was trained with nearly 2500 cutout images of 31$\times$31 pixels from the artefact dataset and real source dataset. The Artefact dataset was synthesised using artefact sources cutout from difference images of the ILMT images in g',r' and i' bands and public data from ZTF \citep{2019PASP..131a8002B}. The real source dataset mainly consisted of point-like sources in g', r' and i' bands in both ILMT and ZTF images. The training dataset was prepared by augmenting the original data with 90$^\circ$, 180$^\circ$ and 270$^\circ$ rotations. \texttt{TensorFlow} library was used to make and train the CNN architecture \citep{10.5555/3026877.3026899}.


\textbf{Transient Classifier:} The third step in the pipeline is the 3-way transient classifier which classifies the detected transients into three categories {\it viz.} orphan candidates, variable source candidates and extragalactic candidates. This is a useful step as it helps in filtering the candidates based on the scientific interests of the user essential for further follow-up. Another major advantage of the 3-way classifier step is that it further reduces the number of alerts that will be needed to be checked and verified manually in cases where alerts for supernovae are needed to be followed up \citep{S_nchez_S_ez_2021, Carrasco_Davis_2021}. This is often desirable as it will enable early time follow-up of supernovae from which important properties of the progenitors may be determined.

The 3-way transient classifier is a CNN which was trained with nearly 1600 cutout images of 102$\times$102 pixels of galaxies, stars and \textit{vacant spaces} to mimic scenarios for extragalactic candidates, variable source candidates and orphan candidates respectively. The CNN algorithm (also referred to as an architecture) operates by looking for certain morphological features in the reference image at the position of detection like presence of a clear host galaxy (for extragalactic candidate) and presence of a point source (for a variable source candidate).   


\section{Preliminary results from the proposed pipeline}
\label{results}

\begin{figure}[t]
\centering
\includegraphics[width=\textwidth]{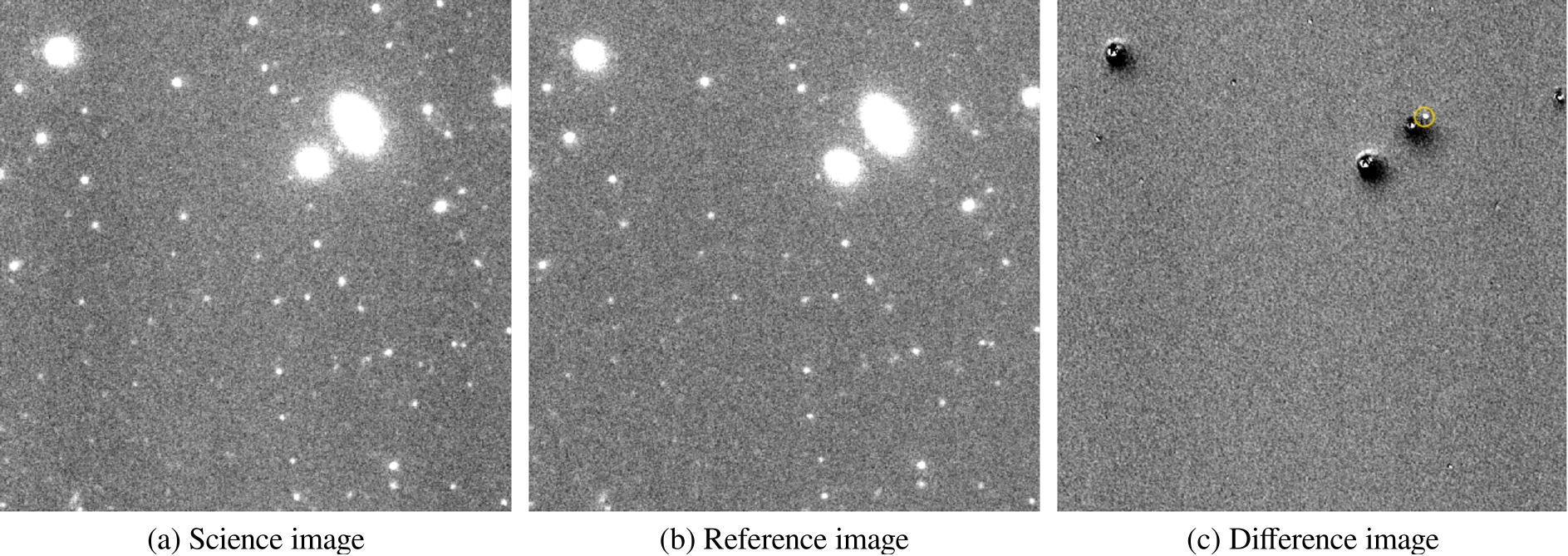}

\begin{minipage}[c]{12cm}
\caption{Image subtraction performed on a 1024$\times$1024 pixel cutout image of an i' band ILMT frame using the difference imaging algorithm. The difference frame shows the presence of the supernova SN 2023af which was recovered by the transient detection step.}
\label{difference_image}
\end{minipage}
\end{figure}

\begin{figure}[t]
\centering
\includegraphics[width=\textwidth]{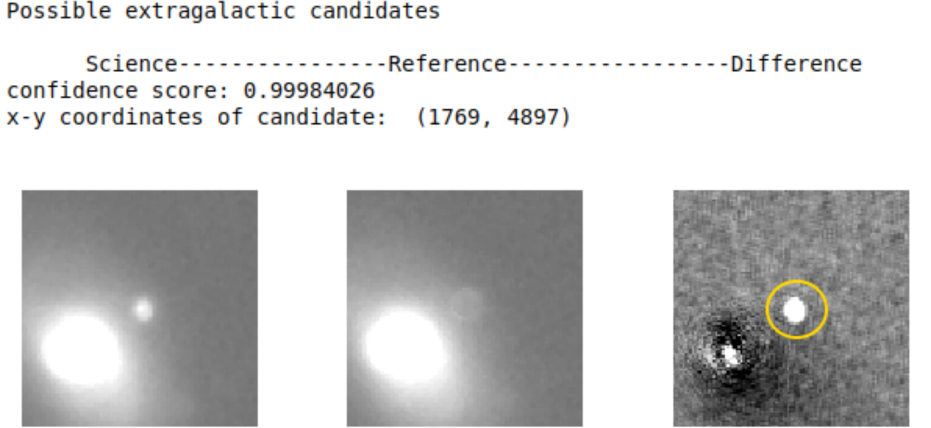}
\smallskip

\begin{minipage}[c]{12cm}
\caption{Classification of SN~2023af (marked with a yellow circle) as an extragalactic candidate by the 3-way classifier.}
\label{fig:my_label}
\end{minipage}
\end{figure}

Preliminary tests were carried out to check the robustness and results of the three primary steps of the pipeline. Fig.\,\ref{difference_image} shows the image subtraction performed on one such ILMT image acquired on 13\textsuperscript{th} March, 2023 in the i' band. The ILMT image observed on 26\textsuperscript{th} March, 2023 was chosen as the reference image. The residual image (rightmost panel in Fig.\,\ref{difference_image}) indicates the presence of a stellar source. This source is identified as SN~2023af (a type II supernova discovered on 2$^{nd}$ January, 2023 by Xingming Observatory Sky Survey (XOSS)) which was incidentally located in the ILMT strip. The reference image for SN~2023af was prepared by artificially removing the supernova from i' band frame of 26\textsuperscript{th} March, 2023 and replacing it with the local background in its place. Fig.\,\ref{fig:my_label} illustrates the classification of the detected SN~2023af as a possible extragalactic candidate with a confidence score of 0.9998 by the transient classifier step. The detection of SN~2023af in the difference image and the classification score indicate the reliability of the pipeline.

\section{Summary and Conclusions}
\label{summary}

The ILMT being a zenith pointing optical survey telescope will be detecting many transients like supernovae. Given the amount of data being acquired each night and the limitations of manpower, an artificial intelligence-based transient detection pipeline is the most suitable strategy.
The various steps of the pipeline have undergone preliminary tests with the ILMT data demonstrating varying degrees of success. It was established during early trials that the various steps of the proposed pipeline were able to detect and classify possible astrophysical sources like asteroids, variable stars, supernovae, etc.
    
As with any machine learning-based application, the transient detection pipeline will require a large amount of data for further training. 
Most of the training done till now was based on data collected in the October--November 2022 cycle of the ILMT in the three SDSS bands. The amount of image data in that cycle was relatively small and hence the trained CNN models might be difficult to generalise for future cycles. Since then, a few changes were made in the telescope itself which might have had some effects on the PSF of the images in the future cycles. As more and more data accumulates, a more extensive training dataset will be constructed and the robustness as well as generality of the pipeline will be tested.
     
\begin{acknowledgments}
The 4m International Liquid Mirror Telescope (ILMT) project results from a collaboration between the Institute of Astrophysics and Geophysics (University of Li\`{e}ge, Belgium), the Universities of British Columbia, Laval, Montreal, Toronto, Victoria and York University, and Aryabhatta Research Institute of observational sciencES (ARIES, India). The authors thank Hitesh Kumar, Himanshu Rawat, Khushal Singh and other observing staff for their assistance at the 4m ILMT.  The team acknowledges the contributions of ARIES's past and present scientific, engineering and administrative members in the realisation of the ILMT project. JS wishes to thank Service Public Wallonie, F.R.S.-FNRS (Belgium) and the University of Li\`{e}ge, Belgium for funding the construction of the ILMT. PH acknowledges financial support from the Natural Sciences and Engineering Research Council of Canada, RGPIN-2019-04369. PH and JS thank ARIES for hospitality during their visits to Devasthal. B.A. acknowledges the Council of Scientific $\&$ Industrial Research (CSIR) fellowship award (09/948(0005)/2020-EMR-I) for this work. M.D. acknowledges Innovation in Science Pursuit for Inspired Research (INSPIRE) fellowship award (DST/INSPIRE Fellowship/2020/IF200251) for this work. T.A. thanks Ministry of Higher Education, Science and Innovations of Uzbekistan (grant FZ-20200929344). The authors also thank the referee for reviewing the manuscript and providing detailed comments.
\end{acknowledgments}

\begin{furtherinformation}

\begin{orcids}
\orcid{0000-0003-1637-267X}{Kuntal}{Misra}
\orcid{0000-0002-7005-1976}{Jean}{Surdej}
\end{orcids}



\begin{authorcontributions}
This work results from a long-term collaboration to which all authors have made significant contributions.

\end{authorcontributions}

\begin{conflictsofinterest}
The authors declare no conflict of interest.
\end{conflictsofinterest}

\end{furtherinformation}

\bibliographystyle{bullsrsl-en}

\bibliography{S11-P03_PranshuK}

\begin{thebibliography}{17}
\providecommand{\natexlab}[1]{#1}
\providecommand{\url}[1]{#1}
\providecommand{\urlprefix}{URL }

\bibitem[{Abadi et~al.(2016)Abadi, Barham, Chen, Chen, Davis, Dean, Devin,
  Ghemawat, Irving, Isard, Kudlur, Levenberg, Monga, Moore, Murray, Steiner,
  Tucker, Vasudevan, Warden, Wicke, Yu and Zheng}]{10.5555/3026877.3026899}
Abadi, M., Barham, P., Chen, J., Chen, Z., Davis, A., Dean, J., Devin, M.,
  Ghemawat, S., Irving, G., Isard, M., Kudlur, M., Levenberg, J., Monga, R.,
  Moore, S., Murray, D.~G., Steiner, B., Tucker, P., Vasudevan, V., Warden, P.,
  Wicke, M., Yu, Y. and Zheng, X. (2016) Tensorflow: A system for large-scale
  machine learning.
\newblock In Proceedings of the 12th USENIX Conference on Operating Systems
  Design and Implementation, OSDI'16, pp. 265--283. USENIX Association, USA.

\bibitem[{Alard and Lupton(1998)}]{Alard_1998}
Alard, C. and Lupton, R.~H. (1998) A method for optimal image subtraction.
\newblock ApJ, 503(1), 325.
\newblock \url{https://doi.org/10.1086/305984}.

\bibitem[{{Bellm} et~al.(2019){Bellm}, {Kulkarni}, {Graham}, {Dekany}, {Smith},
  {Riddle}, {Masci}, {Helou}, {Prince}, {Adams}, {Barbarino}, {Barlow},
  {Bauer}, {Beck}, {Belicki}, {Biswas}, {Blagorodnova}, {Bodewits}, {Bolin},
  {Brinnel}, {Brooke}, {Bue}, {Bulla}, {Burruss}, {Cenko}, {Chang}, {Connolly},
  {Coughlin}, {Cromer}, {Cunningham}, {De}, {Delacroix}, {Desai}, {Duev},
  {Eadie}, {Farnham}, {Feeney}, {Feindt}, {Flynn}, {Franckowiak}, {Frederick},
  {Fremling}, {Gal-Yam}, {Gezari}, {Giomi}, {Goldstein}, {Golkhou}, {Goobar},
  {Groom}, {Hacopians}, {Hale}, {Henning}, {Ho}, {Hover}, {Howell}, {Hung},
  {Huppenkothen}, {Imel}, {Ip}, {Ivezi{\'c}}, {Jackson}, {Jones}, {Juric},
  {Kasliwal}, {Kaspi}, {Kaye}, {Kelley}, {Kowalski}, {Kramer}, {Kupfer},
  {Landry}, {Laher}, {Lee}, {Lin}, {Lin}, {Lunnan}, {Giomi}, {Mahabal}, {Mao},
  {Miller}, {Monkewitz}, {Murphy}, {Ngeow}, {Nordin}, {Nugent}, {Ofek},
  {Patterson}, {Penprase}, {Porter}, {Rauch}, {Rebbapragada}, {Reiley},
  {Rigault}, {Rodriguez}, {van Roestel}, {Rusholme}, {van Santen}, {Schulze},
  {Shupe}, {Singer}, {Soumagnac}, {Stein}, {Surace}, {Sollerman}, {Szkody},
  {Taddia}, {Terek}, {Van Sistine}, {van Velzen}, {Vestrand}, {Walters},
  {Ward}, {Ye}, {Yu}, {Yan} and {Zolkower}}]{2019PASP..131a8002B}
{Bellm}, E.~C., {Kulkarni}, S.~R., {Graham}, M.~J., {Dekany}, R., {Smith},
  R.~M., {Riddle}, R., {Masci}, F.~J., {Helou}, G., {Prince}, T.~A., {Adams},
  S.~M., {Barbarino}, C., {Barlow}, T., {Bauer}, J., {Beck}, R., {Belicki}, J.,
  {Biswas}, R., {Blagorodnova}, N., {Bodewits}, D., {Bolin}, B., {Brinnel}, V.,
  {Brooke}, T., {Bue}, B., {Bulla}, M., {Burruss}, R., {Cenko}, S.~B., {Chang},
  C.-K., {Connolly}, A., {Coughlin}, M., {Cromer}, J., {Cunningham}, V., {De},
  K., {Delacroix}, A., {Desai}, V., {Duev}, D.~A., {Eadie}, G., {Farnham},
  T.~L., {Feeney}, M., {Feindt}, U., {Flynn}, D., {Franckowiak}, A.,
  {Frederick}, S., {Fremling}, C., {Gal-Yam}, A., {Gezari}, S., {Giomi}, M.,
  {Goldstein}, D.~A., {Golkhou}, V.~Z., {Goobar}, A., {Groom}, S., {Hacopians},
  E., {Hale}, D., {Henning}, J., {Ho}, A. Y.~Q., {Hover}, D., {Howell}, J.,
  {Hung}, T., {Huppenkothen}, D., {Imel}, D., {Ip}, W.-H., {Ivezi{\'c}},
  {\v{Z}}., {Jackson}, E., {Jones}, L., {Juric}, M., {Kasliwal}, M.~M.,
  {Kaspi}, S., {Kaye}, S., {Kelley}, M. S.~P., {Kowalski}, M., {Kramer}, E.,
  {Kupfer}, T., {Landry}, W., {Laher}, R.~R., {Lee}, C.-D., {Lin}, H.~W.,
  {Lin}, Z.-Y., {Lunnan}, R., {Giomi}, M., {Mahabal}, A., {Mao}, P., {Miller},
  A.~A., {Monkewitz}, S., {Murphy}, P., {Ngeow}, C.-C., {Nordin}, J., {Nugent},
  P., {Ofek}, E., {Patterson}, M.~T., {Penprase}, B., {Porter}, M., {Rauch},
  L., {Rebbapragada}, U., {Reiley}, D., {Rigault}, M., {Rodriguez}, H., {van
  Roestel}, J., {Rusholme}, B., {van Santen}, J., {Schulze}, S., {Shupe},
  D.~L., {Singer}, L.~P., {Soumagnac}, M.~T., {Stein}, R., {Surace}, J.,
  {Sollerman}, J., {Szkody}, P., {Taddia}, F., {Terek}, S., {Van Sistine}, A.,
  {van Velzen}, S., {Vestrand}, W.~T., {Walters}, R., {Ward}, C., {Ye}, Q.-Z.,
  {Yu}, P.-C., {Yan}, L. and {Zolkower}, J. (2019) {The Zwicky Transient
  Facility: System Overview, Performance, and First Results}.
\newblock PASP, 131(995), 018002.
\newblock \url{https://doi.org/10.1088/1538-3873/aaecbe}.

\bibitem[{Beroiz et~al.(2020)Beroiz, Cabral and S\'{a}nchez}]{astroalign}
Beroiz, M., Cabral, J. and S\'{a}nchez, B. (2020) Astroalign: A python module
  for astronomical image registration.
\newblock A\&C, 32, 100384.
\newblock \url{https://doi.org/10.1016/j.ascom.2020.100384}.

\bibitem[{{Bradley} et~al.(2016){Bradley}, {Sipocz}, {Robitaille}, {Tollerud},
  {Deil}, {Vin{\'\i}cius}, {Barbary}, {G{\"u}nther}, {Bostroem}, {Droettboom},
  {Bray}, {Bratholm}, {Pickering}, {Craig}, {Pascual}, {Greco}, {Donath},
  {Kerzendorf}, {Littlefair}, {Barentsen}, {D'Eugenio} and
  {Weaver}}]{2016ascl.soft09011B}
{Bradley}, L., {Sipocz}, B., {Robitaille}, T., {Tollerud}, E., {Deil}, C.,
  {Vin{\'\i}cius}, Z., {Barbary}, K., {G{\"u}nther}, H.~M., {Bostroem}, A.,
  {Droettboom}, M., {Bray}, E., {Bratholm}, L.~A., {Pickering}, T.~E., {Craig},
  M., {Pascual}, S., {Greco}, J., {Donath}, A., {Kerzendorf}, W., {Littlefair},
  S., {Barentsen}, G., {D'Eugenio}, F. and {Weaver}, B.~A. (2016) {Photutils:
  Photometry tools}.
\newblock ascl soft, record ascl:1609.011.

\bibitem[{Bramich(2008)}]{Bramich_2008}
Bramich, D.~M. (2008) A new algorithm for difference image analysis.
\newblock MNRAS, 386(1), L77--L81.
\newblock \url{https://doi.org/10.1111/j.1745-3933.2008.00464.x}.

\bibitem[{Cabrera-Vives et~al.(2016)Cabrera-Vives, Reyes, F{\"o}rster,
  Est{\'e}vez and Maureira}]{7727206}
Cabrera-Vives, G., Reyes, I., F{\"o}rster, F., Est{\'e}vez, P.~A. and Maureira,
  J.-C. (2016) Supernovae detection by using convolutional neural networks.
\newblock In 2016 IJCNN, pp. 251--258.
\newblock \url{https://doi.org/10.1109/IJCNN.2016.7727206}.

\bibitem[{Carrasco-Davis et~al.(2021)Carrasco-Davis, Reyes, Valenzuela,
  F{\"o}rster, Est{\'{e} }vez, Pignata, Bauer, Reyes, S{\'{a}}nchez-S{\'{a}}ez,
  Cabrera-Vives, Eyheramendy, Catelan, Arredondo, Castillo-Navarrete,
  Rodr{\'{\i}}guez-Mancini, Ruz-Mieres, Moya, Sabatini-Gacit{\'{u}}a,
  Sep{\'{u}}lveda-Cobo, Mahabal, Silva-Farf{\'{a}}n, Camacho-I{\~{n}}iguez and
  Galbany}]{Carrasco_Davis_2021}
Carrasco-Davis, R., Reyes, E., Valenzuela, C., F{\"o}rster, F., Est{\'{e} }vez,
  P.~A., Pignata, G., Bauer, F.~E., Reyes, I., S{\'{a}}nchez-S{\'{a}}ez, P.,
  Cabrera-Vives, G., Eyheramendy, S., Catelan, M., Arredondo, J.,
  Castillo-Navarrete, E., Rodr{\'{\i}}guez-Mancini, D., Ruz-Mieres, D., Moya,
  A., Sabatini-Gacit{\'{u}}a, L., Sep{\'{u}}lveda-Cobo, C., Mahabal, A.~A.,
  Silva-Farf{\'{a}}n, J., Camacho-I{\~{n}}iguez, E. and Galbany, L. (2021)
  Alert classification for the {ALeRCE} broker system: The real-time stamp
  classifier.
\newblock AJ, 162(6), 231.
\newblock \url{https://doi.org/10.3847/1538-3881/ac0ef1}.

\bibitem[{Duev et~al.(2019)Duev, Mahabal, Masci, Graham, Rusholme, Walters,
  Karmarkar, Frederick, Kasliwal, Rebbapragada and Ward}]{Duev_2019}
Duev, D.~A., Mahabal, A., Masci, F.~J., Graham, M.~J., Rusholme, B., Walters,
  R., Karmarkar, I., Frederick, S., Kasliwal, M.~M., Rebbapragada, U. and Ward,
  C. (2019) Real-bogus classification for the zwicky transient facility using
  deep learning.
\newblock MNRAS, 489(3), 3582--3590.
\newblock \url{https://doi.org/10.1093/mnras/stz2357}.

\bibitem[{Huckvale et~al.(2014)Huckvale, Kerins and
  Sale}]{10.1093/mnras/stu835}
Huckvale, L., Kerins, E. and Sale, S.~E. (2014) {Reference image selection for
  difference imaging analysis*}.
\newblock MNRAS, 442(1), 259--272.
\newblock \url{https://doi.org/10.1093/mnras/stu835}.

\bibitem[{Kumar et~al.(2023)Kumar, Dangwal, Rawat, Misra, Negi, Jaiswar,
  Dukiya, Ailawadhi, Hickson and Surdej}]{article_firstlight}
Kumar, B., Dangwal, K., Rawat, H., Misra, K., Negi, V., Jaiswar, M., Dukiya,
  N., Ailawadhi, B., Hickson, P. and Surdej, J. (2023) First light preparations
  of the 4m ilmt.
\newblock JAI, 11.
\newblock \url{https://doi.org/10.1142/S2251171722400037}.

\bibitem[{Kumar et~al.(2018)Kumar, Pandey, Pandey, Hickson, Borra, Anupama and
  Surdej}]{10.1093/mnras/sty298}
Kumar, B., Pandey, K.~L., Pandey, S.~B., Hickson, P., Borra, E.~F., Anupama,
  G.~C. and Surdej, J. (2018) {The zenithal 4-m International Liquid Mirror
  Telescope: a unique facility for supernova studies}.
\newblock MNRAS, 476(2), 2075--2085.
\newblock \url{https://doi.org/10.1093/mnras/sty298}.

\bibitem[{Lecun et~al.(1998)Lecun, Bottou, Bengio and Haffner}]{726791}
Lecun, Y., Bottou, L., Bengio, Y. and Haffner, P. (1998) Gradient-based
  learning applied to document recognition.
\newblock IEEEP, 86(11), 2278--2324.
\newblock \url{https://doi.org/10.1109/5.726791}.

\bibitem[{Mahabal et~al.(2019)Mahabal, Rebbapragada, Walters, Masci,
  Blagorodnova, van Roestel, Ye, Biswas, Burdge, Chang, Duev, Golkhou, Miller,
  Nordin, Ward, Adams, Bellm, Branton, Bue, Cannella, Connolly, Dekany, Feindt,
  Hung, Fortson, Frederick, Fremling, Gezari, Graham, Groom, Kasliwal,
  Kulkarni, Kupfer, Lin, Lintott, Lunnan, Parejko, Prince, Riddle, Rusholme,
  Saunders, Sedaghat, Shupe, Singer, Soumagnac, Szkody, Tachibana, Tirumala,
  van Velzen and Wright}]{Mahabal_2019}
Mahabal, A., Rebbapragada, U., Walters, R., Masci, F.~J., Blagorodnova, N., van
  Roestel, J., Ye, Q.-Z., Biswas, R., Burdge, K., Chang, C.-K., Duev, D.~A.,
  Golkhou, V.~Z., Miller, A.~A., Nordin, J., Ward, C., Adams, S., Bellm, E.~C.,
  Branton, D., Bue, B., Cannella, C., Connolly, A., Dekany, R., Feindt, U.,
  Hung, T., Fortson, L., Frederick, S., Fremling, C., Gezari, S., Graham, M.,
  Groom, S., Kasliwal, M.~M., Kulkarni, S., Kupfer, T., Lin, H.~W., Lintott,
  C., Lunnan, R., Parejko, J., Prince, T.~A., Riddle, R., Rusholme, B.,
  Saunders, N., Sedaghat, N., Shupe, D.~L., Singer, L.~P., Soumagnac, M.~T.,
  Szkody, P., Tachibana, Y., Tirumala, K., van Velzen, S. and Wright, D. (2019)
  Machine learning for the zwicky transient facility.
\newblock PASP, 131(997), 038002.
\newblock \url{https://doi.org/10.1088/1538-3873/aaf3fa}.

\bibitem[{S{\'{a}}nchez-S{\'{a}}ez et~al.(2021)S{\'{a}}nchez-S{\'{a}}ez, Reyes,
  Valenzuela, F{\"o}rster, Eyheramendy, Elorrieta, Bauer, Cabrera-Vives,
  Est{\'{e}}vez, Catelan, Pignata, Huijse, Cicco, Ar{\'{e}}valo,
  Carrasco-Davis, Abril, Kurtev, Borissova, Arredondo, Castillo-Navarrete,
  Rodriguez, Ruz-Mieres, Moya, Sabatini-Gacit{\'{u}}a, Sep{\'{u}}lveda-Cobo and
  Camacho-I{\~{n}}iguez}]{S_nchez_S_ez_2021}
S{\'{a}}nchez-S{\'{a}}ez, P., Reyes, I., Valenzuela, C., F{\"o}rster, F.,
  Eyheramendy, S., Elorrieta, F., Bauer, F.~E., Cabrera-Vives, G.,
  Est{\'{e}}vez, P.~A., Catelan, M., Pignata, G., Huijse, P., Cicco, D.~D.,
  Ar{\'{e}}valo, P., Carrasco-Davis, R., Abril, J., Kurtev, R., Borissova, J.,
  Arredondo, J., Castillo-Navarrete, E., Rodriguez, D., Ruz-Mieres, D., Moya,
  A., Sabatini-Gacit{\'{u}}a, L., Sep{\'{u}}lveda-Cobo, C. and
  Camacho-I{\~{n}}iguez, E. (2021) Alert classification for the {ALeRCE} broker
  system: The light curve classifier.
\newblock AJ, 161(3), 141.
\newblock \url{https://doi.org/10.3847/1538-3881/abd5c1}.

\bibitem[{Sedaghat and Mahabal(2018)}]{10.1093/mnras/sty613}
Sedaghat, N. and Mahabal, A. (2018) {Effective image differencing with
  convolutional neural networks for real-time transient hunting}.
\newblock MNRAS, 476(4), 5365--5376.
\newblock \url{https://doi.org/10.1093/mnras/sty613}.

\bibitem[{Virtanen et~al.(2020)Virtanen, Gommers, Oliphant, Haberland, Reddy,
  Cournapeau, Burovski, Peterson, Weckesser, Bright, van~der Walt, Brett,
  Wilson, Millman, Mayorov, Nelson, Jones, Kern, Larson, Carey, Polat, Feng,
  Moore, VanderPlas, Laxalde, Perktold, Cimrman, Henriksen, Quintero, Harris,
  Archibald, Ribeiro, Pedregosa, van Mulbregt, Vijaykumar, Bardelli, Rothberg,
  Hilboll, Kloeckner, Scopatz, Lee, Rokem, Woods, Fulton, Masson,
  H\"{a}ggstr\"{o}m, Fitzgerald, Nicholson, Hagen, Pasechnik, Olivetti, Martin,
  Wieser, Silva, Lenders, Wilhelm, Young, Price, Ingold, Allen, Lee, Audren,
  Probst, Dietrich, Silterra, Webber, Slavi{\v{c}}, Nothman, Buchner, Kulick,
  Sch\"{o}nberger, de~Miranda~Cardoso, Reimer, Harrington, Rodr{\'{\i}}guez,
  Nunez-Iglesias, Kuczynski, Tritz, Thoma, Newville, K\"{u}mmerer, Bolingbroke,
  Tartre, Pak, Smith, Nowaczyk, Shebanov, Pavlyk, Brodtkorb, Lee, McGibbon,
  Feldbauer, Lewis, Tygier, Sievert, Vigna, Peterson, More, Pudlik, Oshima,
  Pingel, Robitaille, Spura, Jones, Cera, Leslie, Zito, Krauss, Upadhyay,
  Halchenko and and}]{Virtanen_2020}
Virtanen, P., Gommers, R., Oliphant, T.~E., Haberland, M., Reddy, T.,
  Cournapeau, D., Burovski, E., Peterson, P., Weckesser, W., Bright, J.,
  van~der Walt, S.~J., Brett, M., Wilson, J., Millman, K.~J., Mayorov, N.,
  Nelson, A. R.~J., Jones, E., Kern, R., Larson, E., Carey, C.~J., Polat,
  {\.{I}}., Feng, Y., Moore, E.~W., VanderPlas, J., Laxalde, D., Perktold, J.,
  Cimrman, R., Henriksen, I., Quintero, E.~A., Harris, C.~R., Archibald, A.~M.,
  Ribeiro, A.~H., Pedregosa, F., van Mulbregt, P., Vijaykumar, A., Bardelli,
  A.~P., Rothberg, A., Hilboll, A., Kloeckner, A., Scopatz, A., Lee, A., Rokem,
  A., Woods, C.~N., Fulton, C., Masson, C., H\"{a}ggstr\"{o}m, C., Fitzgerald,
  C., Nicholson, D.~A., Hagen, D.~R., Pasechnik, D.~V., Olivetti, E., Martin,
  E., Wieser, E., Silva, F., Lenders, F., Wilhelm, F., Young, G., Price, G.~A.,
  Ingold, G.-L., Allen, G.~E., Lee, G.~R., Audren, H., Probst, I., Dietrich,
  J.~P., Silterra, J., Webber, J.~T., Slavi{\v{c}}, J., Nothman, J., Buchner,
  J., Kulick, J., Sch\"{o}nberger, J.~L., de~Miranda~Cardoso, J.~V., Reimer,
  J., Harrington, J., Rodr{\'{\i}}guez, J. L.~C., Nunez-Iglesias, J.,
  Kuczynski, J., Tritz, K., Thoma, M., Newville, M., K\"{u}mmerer, M.,
  Bolingbroke, M., Tartre, M., Pak, M., Smith, N.~J., Nowaczyk, N., Shebanov,
  N., Pavlyk, O., Brodtkorb, P.~A., Lee, P., McGibbon, R.~T., Feldbauer, R.,
  Lewis, S., Tygier, S., Sievert, S., Vigna, S., Peterson, S., More, S.,
  Pudlik, T., Oshima, T., Pingel, T.~J., Robitaille, T.~P., Spura, T., Jones,
  T.~R., Cera, T., Leslie, T., Zito, T., Krauss, T., Upadhyay, U., Halchenko,
  Y.~O. and and, Y. V.-B. (2020) {SciPy} 1.0: fundamental algorithms for
  scientific computing in python.
\newblock NatMe, 17(3), 261--272.
\newblock \url{https://doi.org/10.1038/s41592-019-0686-2}.

\end{thebibliography}

\end{document}